\documentclass[onecollarge,natbib]{svjour2}
\bibpunct{[}{]}{;}{n}{}{,} 
\smartqed  
\usepackage{graphicx}

\usepackage{amsmath}
\usepackage{latexsym}
\usepackage{amssymb}
\usepackage{dsfont}
\newcommand{\ud}{\mathrm{d}}

\newcommand{\uvec}[1]{\vec{#1}}
\newcommand{\pure}{\text{pure}}
\newcommand{\phys}{\text{phys}}

\journalname{Few-Body Systems}
\begin{document}

\title{Spin structure of the nucleon on the light front}


\author{C\'edric Lorc\'e}


\institute{IFPA, AGO Department, Universit\'e de Li\`ege, Sart-Tilman, 4000 Li\`ege, Belgium \\
              \email{C.Lorce@ulg.ac.be}       
}

\date{Received: date / Accepted: date}

\maketitle

\begin{abstract}
We briefly review the spin structure of the nucleon and show that it is best thought in the light-front formulation. We discuss in particular the longitudinal and transverse spin sum rules, the proper definition of canonical orbital angular momentum and the spin-orbit correlation.
\end{abstract}

\section{Introduction}	

One of the major challenges in hadronic physics is to understand how the nucleon spin arises from the spin and orbital motion of its constituents. Unlike atomic systems, relativistic and non-perturbative effects are essential to understand this spin structure.

The decomposition of the nucleon spin is not unique and is therefore sometimes considered as unphysical. Already at the classical level there exist two definitions (canonical and kinetic) of orbital angular momentum (OAM). While there are often no practical differences between these two definitions, the situation changes in presence of gauge fields, triggering longstanding debates about which definition has to be considered as the fundamental, primary or ``physical'' one. In the context of the nucleon spin decomposition, it has recently been recognized that both definitions are equally interesting, as they are in principle both measurable and reflect complementary aspects of the intricate bound system.

Since measurable quantities are necessarily gauge invariant, recent theoretical works demonstrated that canonical quantities can be made gauge invariant by sacrifying locality or manifest Lorentz covariance. This can be done in infinitely many ways, but only few variants have a clear relation with actual experimental observables. One of the crucial questions now is how to connect the two definitions of OAM to experimental observables. 

Another delicate question is the meaningful separation of spin and OAM. The light-front formalism, with its preferred direction, turns out to provide the most consistent and intuitive picture. Over the last two decades, so many relations and sum rules have been proposed in the literature that they created some sort of confusion. One of the current tasks is to clarify the validity and scope of these relations and sum rules.

In this proceeding, we sketch a portrait of the present situation and present some recent developments. In section \ref{sec2}, we briefly introduce the two families of nucleon spin decompositions. In section \ref{sec3}, we discuss various spin sum rules and relations. In section \ref{sec4}, we define the quark spin-orbit correlation and show how it can be expressed in terms of parton distributions. Finally, we conclude with section \ref{sec5}. For the interested reader, more detailed discussions can be found in the recent reviews~\cite{Leader:2013jra,Wakamatsu:2014zza}.

\section{Canonical and kinetic spin decompositions}\label{sec2}

\begin{figure}
\centering
  \includegraphics[width=8cm]{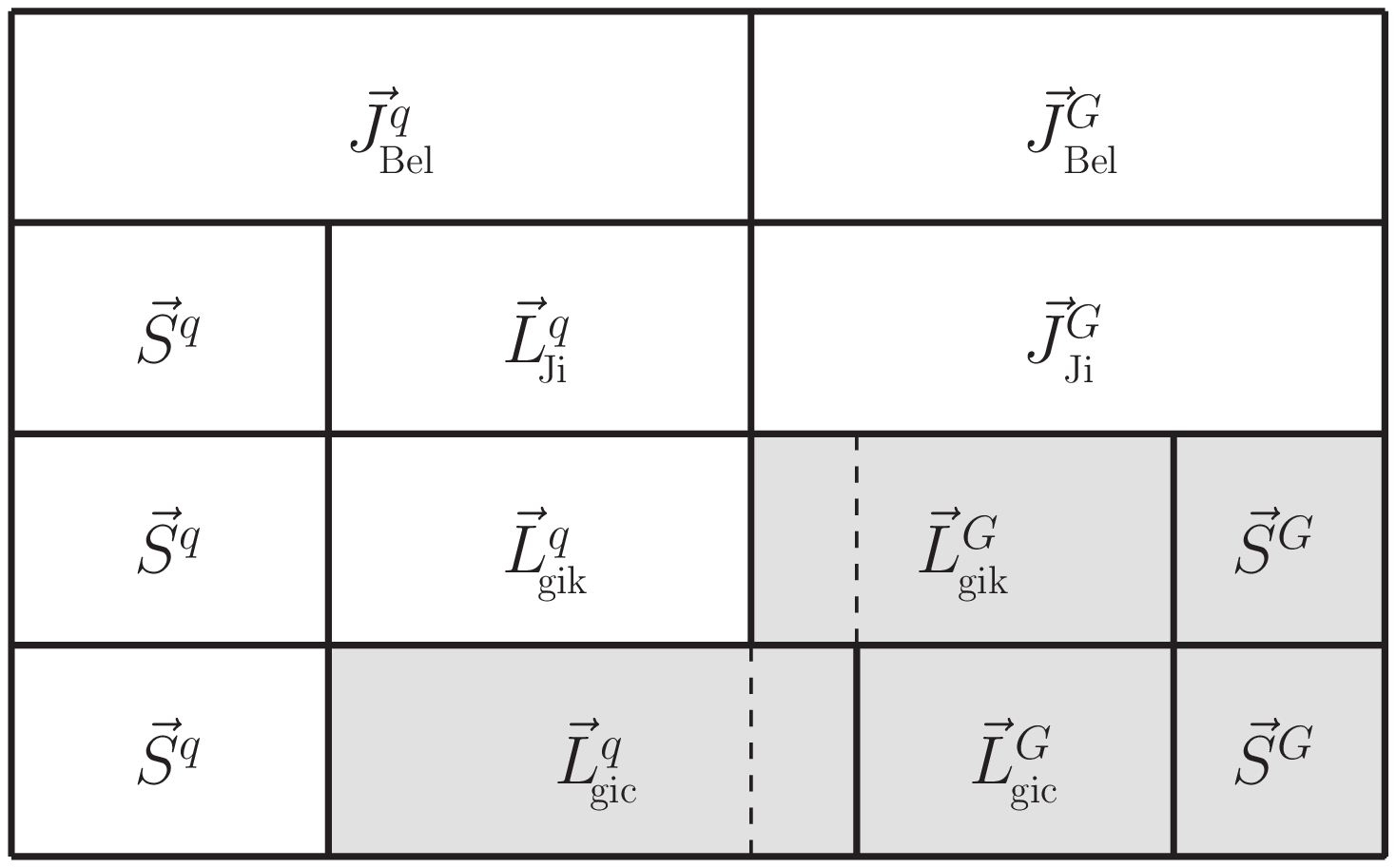}
\caption{The nucleon spin decompositions. See text for more details.}
\label{fig:1}       
\end{figure}

There are essentially two families of nucleon spin decomposition: kinetic and canonical. 
In the kinetic family, one has the Belinfante, Ji and gauge-invariant kinetic (gik) decompositions, depending on how many pieces the total angular momentum is split into, see Fig.~\ref{fig:1}. In the Belinfante decomposition, the total angular momentum is simply decomposed into quark and gluon contributions
\begin{equation}
\uvec J=\uvec J^q_\text{Bel}+\uvec J^G_\text{Bel}.
\end{equation}
Ji decomposed the quark angular momentum into spin and OAM contributions~\cite{Ji:1996ek}
\begin{equation}
\uvec J=\uvec S^q+\uvec L^q_\text{Ji}+\uvec J^G_\text{Ji}\qquad\text{with} \qquad\uvec J^G_\text{Ji}=\uvec J^G_\text{Bel}.
\end{equation}
Finally, decomposing further the gluon angular momentum into spin and OAM gives the gauge-invariant kinetic decomposition discussed by Wakamatsu~\cite{Wakamatsu:2010qj,Wakamatsu:2010cb}
\begin{equation}\label{gik}
\uvec J=\uvec S^q+\uvec L^q_\text{gik}+\uvec S^G+\uvec L^G_\text{gik}\qquad\text{with} \qquad\uvec L^q_\text{gik}=\uvec L^q_\text{Ji}.
\end{equation}
The gauge-invariant canonical (gic) decomposition obtained by Chen \emph{et al.}~\cite{Chen:2008ag,Chen:2009mr}
\begin{equation}\label{gic}
\uvec J=\uvec S^q+\uvec L^q_\text{gic}+\uvec S^G+\uvec L^G_\text{gic}
\end{equation}
can be understood as a gauge-invariant version (or extension) of the Jaffe-Manohar decomposition~\cite{Jaffe:1989jz}, and differs from the gauge-invariant kinetic decomposition in how the total OAM is split into quark and gluon contributions
\begin{equation}
\uvec L=\uvec L^q_\text{gik}+\uvec L^G_\text{gik}=\uvec L^q_\text{gic}+\uvec L^G_\text{gic}.
\end{equation}
A nice physical interpretation of the difference has been proposed in Ref.~\cite{Burkardt:2012sd}.

The operators associated with the various contributions are
\begin{equation}
\begin{aligned}
\uvec S^q&=\int\ud^3x\,\psi^\dag\tfrac{1}{2}\uvec\Sigma\psi, \qquad&\uvec S^G&=\int\ud^3x\,\uvec E^a\times\uvec A^a_\phys,\\
\uvec L^q_\text{gic}&=\int\ud^3x\,\psi^\dag(\uvec x\times i\uvec D_\pure)\psi,\qquad&\uvec L^G_\text{gic}&=-\int\ud^3x\,E^{ai}(\uvec x\times\uvec{\mathcal D}^{ab}_\pure) A^{bi}_\phys,\\
\uvec L^q_\text{gik}&=\int\ud^3x\,\psi^\dag(\uvec x\times i\uvec D)\psi,\qquad&\uvec L^G_\text{gik}&=\uvec L^G_\text{gic}-\int\ud^3x\,(\uvec{\mathcal D}\cdot\uvec E)^a\,\uvec x\times\uvec A^a_\phys,
\end{aligned}
\end{equation}
where the gauge field has been decomposed into two parts $\uvec A=\uvec A^\pure+\uvec A^\phys$, and the pure-gauge covariant derivatives are given by $\uvec D_\pure=-\uvec\nabla-ig\uvec A_\pure$ and $\uvec{\mathcal D}_\pure=-\uvec\nabla-ig[\uvec A_\pure,\quad]$. 

The complete gauge-invariant decompositions~\eqref{gik} and~\eqref{gic} seem to be in contradiction with textbook claims that it is not possible to write down gauge-invariant expressions for gluon spin and OAM contributions. This is actually not the case since textbooks refer only to local expressions, whereas $\uvec A^\pure$ and $\uvec A^\phys$ are non-local expressions of the gauge field~\cite{Lorce:2012rr,Lorce:2012ce}. The pure-gauge part $\uvec A^\pure$ plays essentially the role of a background field~\cite{Lorce:2013gxa,Lorce:2013bja}. Background dependence implies that the split $\uvec A=\uvec A^\pure+\uvec A^\phys$ is accompanied by a new freedom
\begin{equation}\label{Stueckelberg}
\uvec A^\pure\mapsto\uvec A^\pure+\uvec B,\qquad\uvec A^\phys\mapsto\uvec A^\phys-\uvec B,
\end{equation}
referred to as Stueckelberg symmetry~\cite{Lorce:2012rr,Stoilov:2010pv}. The crucial point is that it is the actual experimental conditions that determine the form of the background field to be used~\cite{Lorce:2012rr,Wakamatsu:2014toa}.

\section{Spin sum rules and relations}\label{sec3}

Using the Belinfante-Rosenfeld energy-momentum tensor, Ji obtained the remarkable result that the quark and gluon total kinetic angular momentum can be expressed in terms of twist-2 generalized parton distributions (GPDs)~\cite{Ji:1996ek}
\begin{equation}\label{Jirel}
\langle J^{q,G}\rangle=\tfrac{1}{2}\int\ud x\,x[H_{q,G}(x,0,0)+E_{q,G}(x,0,0)].
\end{equation}
This relation holds for the longitudinal component $J_L=\uvec J\cdot\uvec P/|\uvec P|$ where $\uvec P$ is the nucleon momentum~\cite{Leader:2012md}. By rotational symmetry, it holds also for the transverse component, but only in the nucleon rest frame. Considering the transverse component of the Pauli-Lubanski vector does not prevent frame dependence of the separate quark and gluon contributions~\cite{Leader:2013jra,Leader:2012ar,Hatta:2012jm,Harindranath:2013goa}. 

Subtracting from Eq.~\eqref{Jirel} the longitudinal quark spin contribution given by the isoscalar axial-vector form factor (FF) in the $\overline{MS}$ scheme
\begin{equation}
\langle S^q\rangle=\tfrac{1}{2}\,G^q_A(0),
\end{equation}
one gets the longitudinal quark kinetic OAM 
\begin{equation}\label{OAMeq}
\langle L^q_\text{gik}\rangle=\tfrac{1}{2}\int\ud x\,x[H_{q,G}(x,0,0)+E_{q,G}(x,0,0)]-\tfrac{1}{2}\,G^q_A(0).
\end{equation}
The same quantity is also directly related to a twist-3 GPD~\cite{Penttinen:2000dg,Kiptily:2002nx,Hatta:2012cs}
\begin{equation}
\langle L^q_\text{gik}\rangle=-\int\ud x\,xG^q_2(x,0,0).
\end{equation}

Using the light-front formalism which is particularly suitable for the parton model picture, the most intuitive expression for OAM is as a phase-space integral \cite{Lorce:2011kd,Lorce:2011ni}
\begin{equation}\label{OAMWigner}
\langle L^q(\mathcal W)\rangle=\int\ud x\,\ud^2k_\perp\,\ud^2b_\perp\,(\vec b_\perp\times\vec k_\perp)_z\,\rho^{[\gamma^+]q}_{++}(x,\vec k_\perp,\vec b_\perp;\mathcal W),
\end{equation}
where the relativistic phase-space or Wigner distribution $\rho^{[\gamma^+]q}_{++}(x,\vec k_\perp,\vec b_\perp;\mathcal W)$ can be interpreted as giving the quasi-probability to find an unpolarized quark with momentum $(xP^+,\uvec k_\perp)$ and transverse position $\uvec b_\perp$ inside a longitudinally polarized nucleon. Note that the Euclidean subgroup of the light-front formalism plays here a crucial role in providing a well-defined transverse center of the nucleon~\cite{Soper:1976jc,Burkardt:2000za,Burkardt:2005hp}. The phase-space distributions are related by Fourier transform to the generalized transverse-momentum dependent distributions (GTMDs)~\cite{Meissner:2009ww,Lorce:2011dv,Lorce:2013pza}, leading to the simple relation~\cite{Lorce:2011kd,Hatta:2011ku,Kanazawa:2014nha}
\begin{equation}
\langle L^q(\mathcal W)\rangle=-\int\ud x\,\ud^2k_\perp\,\tfrac{\uvec k^2_\perp}{M^2}\,F^q_{14}(x,0,\uvec k_\perp,\uvec 0_\perp;\mathcal W).
\end{equation}
Depending on the shape of the Wilson line $\mathcal W$, one obtains either kinetic $\langle L^q_\text{gik}\rangle=\langle L^q(\mathcal W_\text{straight})\rangle$ or canonical $\langle L^q_\text{gic}\rangle=\langle L^q(\mathcal W_\text{staple})\rangle$ OAM~\cite{Burkardt:2012sd,Lorce:2012ce,Ji:2012sj}. Unfortunately, it is not known so far how to extract quark GTMDs from actual experiments. They are however in principle calculable on the lattice~\cite{Ji:2013dva}.

Some quark model calculations suggested that the canonical OAM might also be expressed in terms of a transverse-momentum dependent distributions (TMDs) 
\begin{equation}
\langle L_\text{gic}^q\rangle=-\int\ud x\,\ud^2k_\perp\,\tfrac{\uvec k_\perp^2}{2M^2}\,h_{1T}^{\perp q}(x,\uvec k^2_\perp),
\end{equation}
but this relation does not hold in general~\cite{Lorce:2011kn} just like other relations among the TMDs~\cite{Lorce:2011zta}.

\section{Spin-orbit correlation}\label{sec4}

The so-called quark OAM contribution to the nucleon spin corresponds to the correlation between the quark OAM and the nucleon spin. Another interesting quantity is the correlation between the quark spin and OAM which is given by the following operators
\begin{equation}
\begin{aligned}
C^q_\text{gic}&=\int\ud^3x\,\psi^\dag\gamma_5(\uvec x_\perp\times i\uvec D_{\pure,\perp})_z\psi,\\
C^q_\text{gik}&=\int\ud^3x\,\psi^\dag\gamma_5(\uvec x_\perp\times i\uvec D_\perp)_z\psi.
\end{aligned}
\end{equation}
These operators are very similar to the longitudinal quark OAM operators and represent, respectively, the canonical and kinetic versions of the quark spin-orbit correlation~\cite{Lorce:2011kd,Lorce:2014mxa}.

Following a similar approach to Ref.~\cite{Ji:1996ek}, we derived an expression which relates measurable parton distributions to the kinetic version of the quark spin-orbit correlation~\cite{Lorce:2014mxa}
\begin{equation}\label{SOtwist2}
\langle C^q_\text{gik}\rangle=\tfrac{1}{2}\int\ud x\,x\tilde H_q(x,0,0)-\tfrac{1}{2}\,[F^q_1(0)-\tfrac{m_q}{2M_N}\,H^q_1(0)],
\end{equation}
Beside its resemblance with Eq.~\eqref{OAMeq}, the remarkable fact about this expression is that it provides a physical interest in the second moment of the helicity distribution. We also found a corresponding twist-3 GPD relation
\begin{equation}\label{SOtwist3}
\langle C^q_\text{gik}\rangle=-\int\ud x\,x[\tilde G^q_2(x,0,0)+2\tilde G^q_4(x,0,0)].
\end{equation}

Like the quark OAM, the most intuitive expression for the quark spin-orbit correlation is as a phase-space integral \cite{Lorce:2011kd,Lorce:2014mxa}
\begin{equation}\label{OAMWigner}
\langle C^q(\mathcal W)\rangle=\int\ud x\,\ud^2k_\perp\,\ud^2b_\perp\,(\vec b_\perp\times\vec k_\perp)_z\,\rho^{[\gamma^+\gamma_5]q}_{++}(x,\vec k_\perp,\vec b_\perp;\mathcal W),
\end{equation}
where the phase-space distribution $\rho^{[\gamma^+\gamma_5]q}_{++}(x,\vec k_\perp,\vec b_\perp;\mathcal W)$ can be interpreted as giving the difference between the quasi-probabilistic distributions of quarks with polarization aligned and antialigned with respect to the longitudinal direction. The simple relation in terms of GTMDs is~~\cite{Lorce:2011kd,Kanazawa:2014nha,Lorce:2014mxa}
\begin{equation}
\langle C^q(\mathcal W)\rangle=\int\ud x\,\ud^2k_\perp\,\tfrac{\uvec k^2_\perp}{M^2}\,G^q_{11}(x,0,\uvec k_\perp,\uvec 0_\perp;\mathcal W).
\end{equation}
Once again, depending on the shape of $\mathcal W$, one obtains either kinetic $\langle C^q_\text{gik}\rangle=\langle C^q(\mathcal W_\text{straight})\rangle$ or canonical $\langle C^q_\text{gic}\rangle=\langle C^q(\mathcal W_\text{staple})\rangle$ spin-orbit correlation~\cite{Lorce:2014mxa}.

Interestingly, since $F^u_1(0)=2$ and $F^d_1(0)=1$ and since the tensor FF $H^q_1(0)$ can safely be neglected because of the mass ratio $m_{u,d}/4M_N\sim 10^{-3}$, the essential input we need is the second moment of the quark helicity distribution
\begin{equation}
\int_{-1}^1\ud x\,x\tilde H_q(x,0,0)=\int_0^1\ud x\,x[\Delta q(x)-\Delta\overline q(x)].
\end{equation}
Contrary to the lowest moment $\int_{-1}^1\ud x\,\tilde H_q(x,0,0)=\int_0^1\ud x\,[\Delta q(x)+\Delta\overline q(x)]$, this second moment cannot be extracted from deep-inelastic scattering (DIS) polarized data without extra assumptions about the polarized sea-quark distributions. However, by combining inclusive and semi-inclusive deep-inelastic scattering, separate quark and antiquark contributions can be extracted~\cite{Leader:2010rb}
\begin{equation}
\int_{-1}^1\ud x\,x\tilde H_u(x,0,0)\approx 0.19,\qquad \int_{-1}^1\ud x\,x\tilde H_d(x,0,0)\approx -0.06,
\end{equation}
at the scale $\mu^2=1$ GeV$^2$, leading to the values $C^u_z\approx -0.9$ and $C^d_z\approx -0.53$. These values seem consistent with recent Lattice calculations by the LHPC collaboration~\cite{Bratt:2010jn}, see table~\ref{Modelresults}.

\begin{table}[t]
\caption{Comparison between the lowest two axial moments for $u$ and $d$ quarks as predicted by the naive quark model (NQM), the light-front constituent quark model (LFCQM) and the light-front chiral quark-soliton model (LF$\chi$QSM) at the scale $\mu^2\sim 0.26$ GeV$^2$, with the corresponding values obtained from the LSS fit to experimental data at $\mu^2=1$ GeV$^2$ and Lattice calculations at $\mu^2=4$ GeV$^2$ and pion mass $m_\pi=293$ MeV.}
\centering\label{Modelresults}
\begin{tabular}{ccccc}\hline\noalign{\smallskip}
Model~\cite{Lorce:2011dv}&$\int^1_{-1}\ud x\,\tilde H_u(x,0,0)$&$\int^1_{-1}\ud x\,\tilde H_d(x,0,0)$&$\int^1_{-1}\ud x\,x\tilde H_u(x,0,0)$&$\int^1_{-1}\ud x\,x\tilde H_d(x,0,0)$\\
\hline\noalign{\smallskip}
NQM&$4/3$&$-1/3$&$4/9$&$-1/9$\\
LFCQM&$0.995$&$-0.249$&$0.345$&$-0.086$\\
LF$\chi$QSM&$1.148$&$-0.287$&$0.392$&$-0.098$\\
\hline\noalign{\smallskip}
LSS~\cite{Leader:2010rb}&$0.82$&$-0.45$&$\approx 0.19$&$\approx -0.06$\\
Lattice~\cite{Bratt:2010jn}&$0.82(7)$&$-0.41(7)$&$\approx 0.20$&$\approx -0.05$\\
\noalign{\smallskip}\hline
\end{tabular}
\end{table}

Noting that the second moment of the quark helicity distribution is a valence-like quantity with tamed low-$x$ region, one may expect phenomenological quark model predictions to be more accurate than for the lowest moment. In table~\ref{Modelresults} we provide the first two moments of the up and down quark helicity distributions obtained within the naive quark model (NQM), the light-front constituent quark model (LFCQM)~\cite{Boffi:2002yy,Boffi:2003yj,Pasquini:2005dk,Pasquini:2006iv,Pasquini:2008ax} and the light-front chiral quark-soliton model (LF$\chi$QSM)~\cite{Lorce:2006nq,Lorce:2007as,Lorce:2007fa} at the scale $\mu^2\sim 0.26$ GeV$^2$. From these estimates, we expect a negative quark spin-orbit $C^q_z$ for both quark flavors ($C^u_z\approx -0.8$ and $C^d_z\approx -0.55$), meaning that the quark spin and kinetic OAM are expected to be, in average, antiparallel. On the contrary, the canonical version of the quark spin-orbit correlation turns out to be positive in the models~\cite{Lorce:2011kd}, displaying the importance of the quark-gluon interaction.

\section{Conclusion}\label{sec5}

There are essentially two types or families of nucleon spin decompositions: the canonical one and the kinetic one. It has recently been recognized that both are in principle measurable. The crucial piece which is currently missing is the contribution coming from the quark and gluon orbital angular momentum. Many relations and sum rules have been proposed, but few turned out to be of practical significance. The light-front formalism is the best suited for describing and interpreting the high-energy scattering experiments involving nucleons. It is therefore hardly surprizing that it gives the proper formulation for decomposing in a meaningful way the nucleon spin. Finally, the quark spin-orbit correlation is a new independent quantity characterizing the nucleon spin structure. Like the quark orbital angular momentum, this information can be extracted from measurable parton distribution.

\begin{acknowledgements}
I benefited a lot from many discussions and collaborations with E. Leader, B. Pasquini and  M. Wakamatsu. This work was supported by the Belgian Fund F.R.S.-FNRS \emph{via} the contract of Charg\'e de Recherches.
\end{acknowledgements}


\end{document}